\documentclass[sigconf, authordraft=false]{acmart}

\usepackage{booktabs} 

\usepackage{subcaption}

\usepackage[ruled]{algorithm2e} 

\SetAlFnt{\small}
\SetAlCapFnt{\small}
\SetAlCapNameFnt{\small}
\SetAlCapHSkip{0pt}
\IncMargin{-\parindent}

\usepackage{listings}
\usepackage{enumitem}
\usepackage{siunitx}
\usepackage{multirow}

\setlist[enumerate,1]{%
  label=\arabic*.,
}

\newlist{inlinelist}{enumerate*}{1}
\setlist*[inlinelist,1]{%
  label=(\roman*),
}

\setcopyright{rightsretained}

\begin{document}
\title{PCOT: Cache Oblivious Tiling
of Polyhedral Programs}
\author{Waruna Ranasinghe}
\author{Nirmal Prajapati}
\affiliation{%
  \institution{Colorado State University}
  \department{Department of Computer Science}
  \city{Fort Collins}
  \state{CO}
  \postcode{80523}
  \country{USA}}
\author{Tomofumi Yuki}
\affiliation{%
  \institution{INRIA}
  \city{Rennes}
  \country{France}}
\author{Sanjay Rajopadhye}
\affiliation{%
  \institution{Colorado State University}
  \department{Department of Computer Science}
  \city{Fort Collins}
  \state{CO}
  \postcode{80523}
  \country{USA}}
\begin{abstract}
  This paper studies two variants of tiling: iteration space tiling (or loop
  blocking) and cache oblivious methods that recursively split the iteration
  space with divide-and-conquer. The key question to answer is when
  we should be using one over the other. The answer to this question is complicated
  for modern architecture due to a number of reasons.

  In this paper, we present a detailed empirical study to answer this question for a range of kernels
  that fit the polyhedral model. Our study is based on a generalized cache oblivious code generator that
  support this class, which is a superset of those supported by existing tools.
  The conclusion is that cache oblivious code is most useful when the aim is to
  have reduced off-chip memory accesses, e.g., lower energy,
  albeit certain situations that diminish its effectiveness exist. 
\end{abstract}
%
%

%
%


%
%




\maketitle


\section{Introduction}
\label{sec:introduction}

Modern multicore processors are complicated, and programming them is a
challenge, especially when seeking the best performance.  Many different, and
often conflicting factors need to be optimized, notably, parallelism and data
locality, and that too, at multiple levels of the memory/processor hierarchy
(vectors-registers cores-caches).  Indeed, in the exascale era, the very
notion of ``performance'' may refer to execution time (i.e., speed) or to
energy (product of the average power and time), or even the energy-delay
product.

Iteration space tiling~\cite{irigoin-popl88, Wol87, Wolf91tiling} (variously
called loop blocking~\cite{schreiber-TR90} or
partitioning~\cite{bu-deprettere, darte-partitioning91, teich-thiele93}) is a
critical transformation, used for multiple objectives: balancing granularity
of communication to computation across nodes in a distributed machine,
improving data locality on a single node, controlling locality and parallelism
among multiple cores on a node, and, at the finest grain, exploiting
vectorization while avoiding register pressure on a single core.  It is an
essential strategy used by compilers and automatic parallelizers.


Another approach to optimizing memory transfers is provided by cache oblivious
algorithms~\cite{prokop-thesis99, frigo-etal-focs99}.  They come with elegant
theoretical bounds on the number of cache misses, and are claimed to require
less tuning than iteration space tiling.  When applied to specific computation
patterns like ``stencils,'' cache-oblivious the strategy may also be viewed as
tiling, where a tile is recursively split into smaller ones through the
divide-and-conquer strategy~\cite{frigo-strumpen-ics05}.

The fundamental difference between the two approaches is the \emph{schedule},
or temporal order in which tiles are executed. A program transformed with
(single-level) iteration space tiling visits the tiles in the lexicographic
order.  Cache oblivious methods visit (leaf) tiles in a different order,
yielding better locality---this can also be viewed as hierarchical
tiling~\cite{carter1995hierarchical} where the number of levels depends on the
problem size.

The practical impact of this difference is not fully understood due to many
complicating factors.  The theoretical results for cache oblivious methods
assume fully associative caches, and therefore do not fully account for
conflict misses.  Hardware prefetching plays an important role in modern
architecture, and it also has significant impact on the choice of tile
sizes~\cite{mehta2016turbotiling}.  Most importantly, existing tools for cache
oblivious code generation are domain specific: Pochoir targets Jacobi-style
stencils~\cite{Tang2011} and AutoGen specializes on dynamic
programming~\cite{autogen-ppopp16}.  This makes an ``apples-to-apples''
comparison of the two techniques difficult because of the mismatch between the
classes of programs where the application of the two techniques are automated.

Our objective is to understand the difference between iteration space tiling
and cache oblivious strategy on modern architecture.  Our study emphasizes the
influence of tile sizes (or base case thresholds) that strongly impact the
performance of both approaches.  To do this we generalized cache oblivious
code generation to all computations that fit the polyhedral model, i.e.,
(imperfect) loop nests with affine bounds and array
accesses~\cite{feautrier2011polyhedral, sanjay-fsttcs86, quinton-jvsp89,
  quinton-sanjay-tf, feautrier91, feautrier92a, feautrier92b}.  This class of
programs include, as a \emph{proper} subset, all the inputs handled by Pochoir
and Autogen, and many others such as dense linear algebra algorithms and
Gauss-Seidel style stencils.  Our study reveals the following:
\begin{itemize}\itemsep 0mm
\item The absolute number of cache misses, as well as the variability of this
  number as a function of tile sizes, are both significantly lower with cache
  oblivious codes.  This is consistent with theeory.
\item Both approaches show similar variability in speed,
  implying that the tuning effort for both methods is similar.  The main
  reason is that speed does not depend on cache behavior alone.  The leaf tile
  sizes influence other aspects such as recursion overhead, register
  allocation, vectorization, and prefetcher effectiveness.
\item The additional levels of tiling in cache oblivious codes do not
  contribute to speed, especially for polyhedral programs.  Once there the
  code is compute-bound at a certain level of cache, additional locality in
  slower caches does not improve speed.
\item The benefit of cache oblivious approaches diminish on architectures with
  low set-associativity of caches, and with prefetching.  The tile execution
  order of cache oblivious codes can increase conflict misses with high
  dimensional data.  Hardware prefetching may favor large tile sizes in the
  innermost dimension, if speed is the primary objective, where the leaf tile
  sizes are already targeting the LLC (Last Level Cache).
\end{itemize}
These results are, in part, confirmation of the work by Yotov et
al.~\cite{yotov2007experimental}, who focused on sequential executions of
matrix computations.  We extended the study to a broader range of kernels, and
to parallel execution.  The main takeaway is that if the main interest is in
speed, then cache oblivious codes are unlikely to help.  However, if the
goal is to reduce off-chip memory traffic, e.g., for better energy
efficiency, then cache oblivious codes give this for ``free:'' no additional
tuning is required beyond that required for speed.  Even then, its
effectiveness is constrained by various aspects of the code and the platform,
and care must be taken to select the appropriate strategy.  We provide a
generalized cache oblivious code generator to offer this option to any
polyhedral program, and give insights on when it will be effective.

\section{Motivations for Empirical Study}
\label{sec:motivations}
The key question that we try to answer is when and why we should use standard
iteration space tiling over cache oblivious tiling.  The two approaches
perform similar partitioning of the iteration space, but the schedules given
to the partitions are different.  Theoretically, cache oblivious code seems to
have advantages over iteration space tiling.  However, many factors complicate
the actual performance, which made our initial experiments difficult to
interpret.  In this section, we describe the obstacles between the theory and
practice we have identified.

We use Single-Level Tiling (SLT) for iteration space tiling, and Cache
Oblivious Tiling (COT) for cache oblivious techniques in this
paper, which are further described in Section~\ref{sec:background}.

\paragraph{Recursion Overhead} This is a well-known overhead of
COT~\cite{yotov2007experimental}.  The recursion introduces overheads, such as
function call overhead, and increased register pressure.  Furthemore, the
functions force inter-procedural analysis/optimization, known to be more
difficult for compilers well.  Thus, the leaf tiles must be ``sufficiently
large'' to avoid excessive overhead due to the recursion.

 \paragraph{Recursive Split Constraints the Tile Sizes} In typical cache
 oblivious algorithms, the problem is recursively split into halves in each
 dimension. This is in fact a rather coarse-grained exploration of the
 hierarchical partitioning of the iteration space. For instance, if the
 current problem size is $B^3$, then the next sub-problem would be
 $(\frac{B}{2})^3$.  If the best problem size for utilizing a level of cache
 is $(B-x)^3$ where $x\ll \frac{B}{2}$ then the subproblems due to
 divide-and-conquer will not match the best.  This is another factor that
 necessitates fine tuning of leaf tile sizes even for COT, since the utilization
 rate of L1 cache has strong impact on performance.  


\paragraph{COT has more Conflict Misses} The divide-and-conquer execution
order may negatively affect cache interference, especially with high
dimensional data.  This happens when the memory is allocated such that the
accesses are contiguous along some direction in the iteration space (typically
along innermost canonical axis).  With lexicographic order of execution, this
contiguity is largely preserved in the tiled execution.  However,
divide-and-conquer executes neighboring tiles in all dimensions, and many of
those tiles access some distant location in memory.  In contrast to accessing
contiguous regions of memory, accessing various segments of the memory
increases the chances of conflicts.

\paragraph{Hardware Prefetching}  Modern architectures are equipped with
hardware prefetchers that can bring data to the L1 cache. When
having sufficient locality at L2 or LLC makes the program compute-bound, then
the latency to L2/LLC can be hidden by the prefetcher. For such programs, it is
unnecessary to tile for the fastest cache, and larger tiles targeting slower
caches improve performance by maximizing prefetcher
effectiveness~\cite{mehta2016turbotiling}. When the primary objective is speed,
the leaf tiles for COT should also be large, which negates the benefit of
divide-and-conquer, as the leafs are already targeting slower caches.
Prefetching have little impact on parallel executions, since prefetching is
bandwidth limited. When multiple cores try to prefetch at the same time,
the bandwidth limit is quickly reached, and the latency hiding effect is
lost. Furthermore, smaller tile sizes are better for parallel execution for
load balancing  reasons.

These factors limit the effectiveness of COT in various ways and are also
closely tied to the characteristics of the computation. Our empirical study
illustrate the impact of these factors on polyhedral computations.

 
\section{Background}
\label{sec:background}

In this section, we present necessary background on tiling, cache oblivious
divide-and-conquer execution, and define the terminology used in this paper.

\subsection{Tiling}
Tiling is a well-known loop transformation for partitioning computations into
smaller, atomic (all inputs to a tile can be computed before its execution),
units called tiles~\cite{irigoin-popl88, Wolf91tiling}. The partitioning into
tiles improves data locality by altering the execution order of the operations.
Tiling also exposes coarse-grained parallelism making it the core transformation
for polyhedral automatic parallelizers such as Pluto~\cite{uday-pldi08}.

 The natural legality condition of tiling is that the dependences across tiles do not
create a cycle.  In compilers, this condition is typically expressed as fully
permutability (i.e., dependences are non-negative direction vectors), which is
a sufficient condition. In the rest of this paper, we assume that the programs
have been transformed to expose loop nests that satisfy this condition.  For
polyhedral programs, scheduling techniques to expose such loop nests are
available~\cite{uday-pldi08}.

\subsection{Cache Oblivious Tiling}
Cache oblivious algorithms~\cite{prokop-thesis99, frigo-etal-focs99} are based
on recursive formulation into smaller subproblems (divide-and-conquer). The
main argument is that as the problem sizes are recursively made smaller, a
subproblem is going to fit on some level of the memory hierarchy that may be
caches, main memory, etc. This class of algorithms is expected to take
advantage of all memory hierarchies through this strategy.

Cache Oblivious Tiling is a specialization of such algorithms based on tiling.
Tiles after a level of tiling can be tiled again with smaller tile
sizes to realize the divide-and-conquer execution pattern. COT may be viewed as
hierarchical tiling, except that the number of tiling levels are determined at
run-time through divide-and-conquer~\cite{carter1995hierarchical}. 

\begin{figure}
\centering
  \begin{subfigure}{0.2\textwidth}
    \includegraphics[width=\columnwidth]{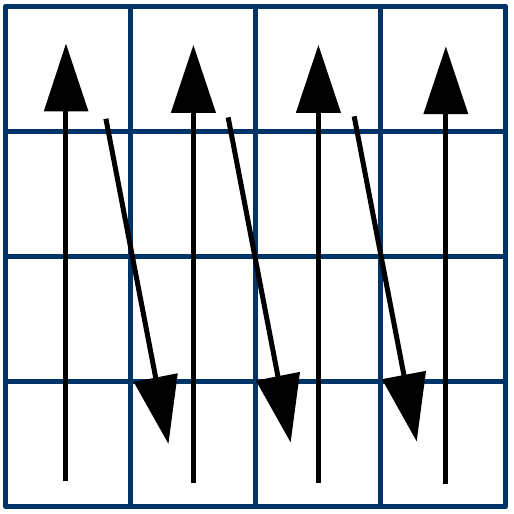}
    \caption{Single-Level Tiling \label{fig:wavefront-tile-order}}
  \end{subfigure}
  \begin{subfigure}{0.2\textwidth}
    \includegraphics[width=\columnwidth]{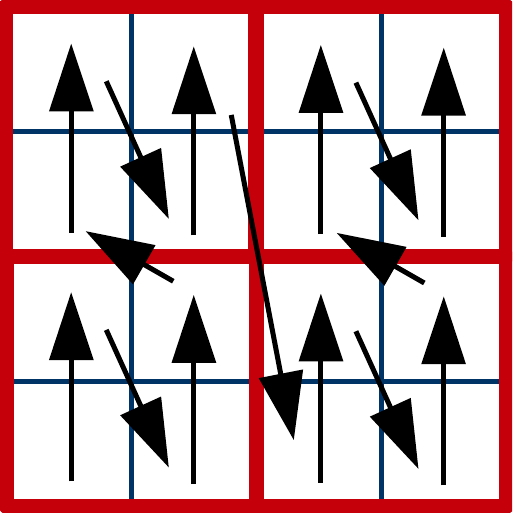}
    \caption{Two-Level Tiling \label{fig:cot-tile-order}}
  \end{subfigure}
  \caption{\label{fig:tile-order}Execution order of the smallest tiles under single- and two-level
tiling. Two-level tiling may be viewed as Cache Oblivious Tiling where the
recursion reached the base case after one recursive step. With hierarchical tiling,
neighboring tiles form a larger tile that increase intra-tile reuse.
\vspace{-0.2cm}
}
\end{figure}

The key effect of multi-level tiling is to change the execution order of the
tiles. As illustrated in Figure~\ref{fig:tile-order}, the grouping of smaller
tiles; forming larger tiles increasing intra-tile reuse; is what accounts for
better cache utilization.

\subsection{Terminology}\label{sec:terminology}
We introduce a few terms, in addition to COT in the above.

\paragraph{Single-Level Tiling} SLT is when loop tiling is applied once to
improve data locality with respect to a level of cache. The reason we restrict
to SLT is because SLT and COT expose the same number of tuning parameters. It
is known that even cache oblivious algorithms require the performance of the
leaf subproblems to be tuned to have good performance. The tuning effort
required is similar to SLT with the same number of tuning parameters, which
would not be the case if you apply multi-level tiling that multiplies the
number of tuning parameters.  

\paragraph{Tile Size} We use tile size interchangeably with the base case
threshold (and leaf tile size) in COT. Whenever tile sizes are discussed for
COT, it refers to the size of the leaf tile sizes, which is its tuning parameter. 

\paragraph{Off-Chip Accesses} OCA are all load accesses that read from the main
memory.  This includes the accesses due to Last-Level Cache misses due to load
operations, as well as those arising from prefetching instructions.

\section{Code Generation of PCOT}
\label{sec:pcot}


In this section, we describe our generalization of the cache oblivious code
generation to polyhedal programs: Polyhedral Cache Oblivious Tiling (PCOT).

\subsection{Approach Overview}
The input is any polyhedral loop nest that is fully permutable and hence
tiling it with hyper-rectangular tiles is a legal transformation.  We first
(in Section~\ref{sec:perfect}) describe the case for tiling all dimensions of a
perfect loop nest.  Other cases can be handled with additional pre-processing,
which we describe in Section~\ref{sec:codegen_ext}.

\begin{figure*}
  \centering
    \includegraphics[width=0.98\textwidth]{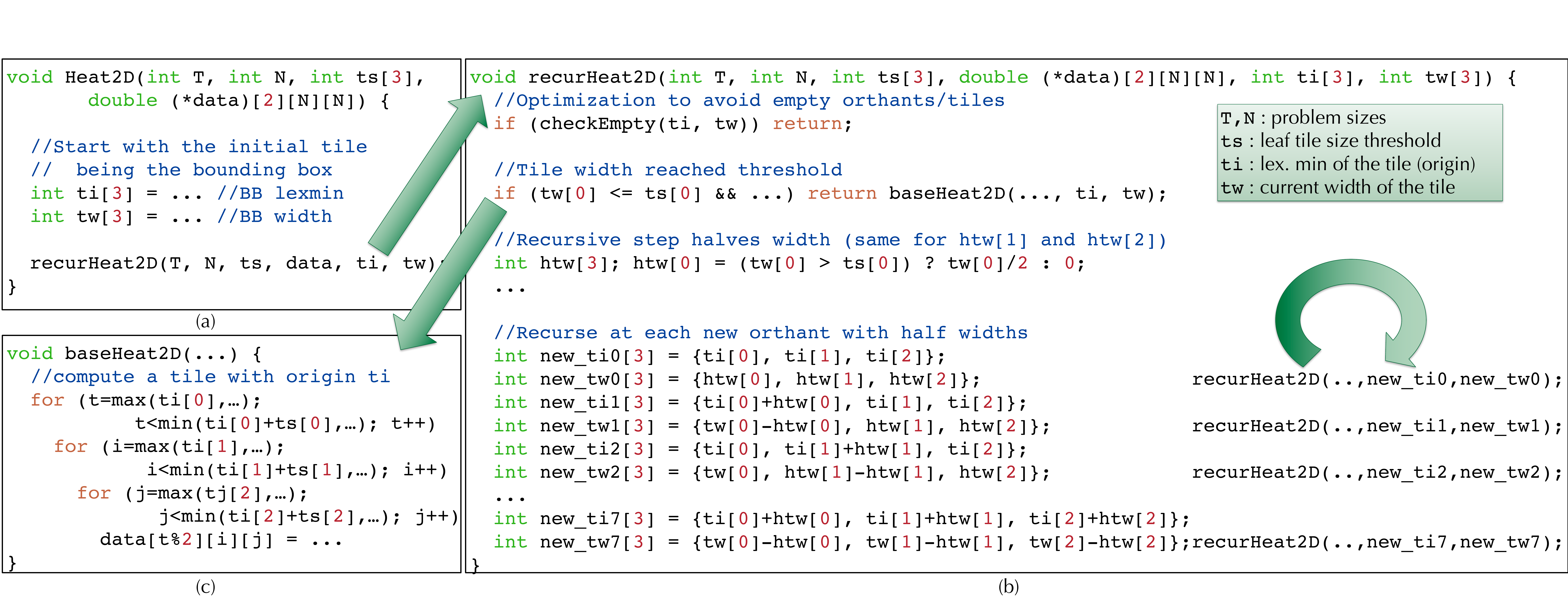}
    \caption{The structure of generated code for Heat2D stencils, which has
      loop depth $d=3$. (a) The tileable loop nest is replaced by a call to
      start the recursion. The input is the bounding box of the loop nest. (b)
      Structure of the recursive function. The input bounding box is split
      into $2^d$ new orthants by dividing each dimention in half, and the same
      function is recursively called for each new orthant. When the orthant
      reaches the input tile size, the recursion is terminated by a call to
      the base function. (c) Code for the base function that
      performs the computation of a tile in lexicographic order.  }
  \label{fig:sample_code}
\end{figure*}

Figure~\ref{fig:sample_code} illustrates the structure of our generated code.
The input loop nest is replaced by a call to start the recursion as shown in 
Figure~\ref{fig:sample_code}a. The computation of the bounding box from loop nests 
are discussed in Section~\ref{sec:computingBB}.

\subsection{Codegen for Perfect Loop Nests}
\label{sec:perfect}

Given a perfectly nested loop with all $d$ dimensions tilable, we seek to
generate,
\begin{inlinelist}
	\item a call to recursive function to start the recursion,
	\item a recursive function, and
	\item a base function.
\end{inlinelist}  
The recursive function takes origin and the size of the orthant as inputs,
which are $d$ dimensional vectors. For the initial call to the recursive
function, the origin and the orthant size correspond to the bounding box of
the input loop nest.  Bounding box of a domain is the smallest
hyper-rectangular shaped domain which encloses the given domain.

The recursive function visits the iteration domain in divide-and-conquer
order.  It recursively divides iteration domain into orthants until they are
smaller than an input parameter.
The call to the base function is wrapped by a condition to check whether the
size of the current orthant is less than or equal to the base case threshold. 

The orthants are visited sequentially in the lexicographic order.  For
parallel execution the tasks are executed with wave-front parallelism.  We use
the OpenMP tasks for parallel execution, where each recursive function call is
annotated with {\tt omp task} pragma, and the wave-front time boundaries are
synchronized with {\tt omp taskwait}.

The base function visits all points in the intersection of leaf orthant and
the input loop nest iteratively in lexicographical order.  The loop nest in
the base function is identical to the loop nest for a tile in SLT with
parametric tile sizes we use.  One can use any one of the available
parametrically tiled code generators~\cite{sanjay-lcpc2009,
  sanjay-kim-dtilingTR-2010,baskaran-etal-cgo10,iooss:hal2015} to generate
parametrically tiled loops and then extract the point loops.

\subsection{Optimizations}
\label{sec:codegen:opt}
We implement a number of optimizations to improve the speed of code.  In this
section we explain two important optimizations to exit the recursion early.

\subsubsection{Early Exit for Zero Orthants}

The zero orthants surface when we have tile sizes that are cubic bounding box with
hyper-rectangular tiles and vise versa. In this case, orthant size along all
the dimensions reaches the leaf size in different levels of the recursion. But
still our code generator generates $2^d$ sequence of recursive calls where
size some of the new orthants may be zero along dimensions where the input
orthant size is already smaller than the leaf threshold.  This is a simple
optimization where we check whether a width along anyone of the dimensions of
the orthant is zero. If it is zero then exit the recursion.

\subsubsection{Early Exit for Empty Orthants}
Second optimization is due to the fact we visit the bounding box of the input
loop nest instead of the actual domain. Some kernels operate on
non-hyper-rectangular domains(i.e, Cholesky  operates on triangular domains)
and some kernels need iteration space skewing to enable hyper-rectangular
tiling (i.e., Heat2D). If the actual domain is not a hyper-rectangle then the
bounding box will have points where no computations are defined.  This may
lead to orthants outside of the original iteration space, analogous to empty
tiles in iteration space tiling. 

For example, a loop nest whose iteration space is triangular, the boundingbox
is a rectangle where the half of the points has no computation defined. When
we visit this rectangular box in divide-and-conquer order, we will end up
visiting empty orthants and we want to identify these orthants.  We generate
conditions to check whether all vertices of the orthant remain outside of the
original iteration space using \emph{isl} library~\cite{verdoolaege2010isl}.


The \texttt{checkEmpty} method at the top of the
Figure~\ref{fig:sample_code}c implements both optimizations.  Without them,
the code produces correct answers but visits many base cases which are empty.
The recursion ends either when \texttt{checkEmpty} method returns true or when
the orthant reaches its input tile size.

\subsection{Handling Imperfect Loop Nests}
\label{sec:codegen_ext}
The discussion so far assumed perfectly nested loops as inputs.  We now extend
this to imperfect loop nests, and loop nests where subset of the $d$
dimensions are marked as tilable.

The input imperfect loop nests are converted to perfect loop nests with a
pre-processing.  This is called the embedding transformation that are used to
handle imperfectly nested loops in parametric tiled code
generation~\cite{sanjay-kim-dtilingTR-2010}. 
It involves, bringing all the statements into the same loop depth by adding
loops with one iteration as necessary. Then affine guards are added to
eliminate sequence of inner loops, which lead to ``perfect loop nest with
affine guards''.


When a subset of the loops are marked as tilable, we first extract the marked
band of loops parameterized by both program input parameters as well
as untiled outer loop iterators. We apply the techniques we have discussed so
far to generate code for the extracted loop nest. In this case, the function
call to start the recursion is added as the body of outer untiled loop nest.
The inner untiled loop nests are added as the body of the point loop in the
base function.

\subsection{Computing the Bounding Box}
\label{sec:computingBB}
The bounding box is a hyper-rectangle containing the iteration space of the
loop nest. It is computed by eliminating the outer loop indices from the loop
bound expressions.  There can be infinitely many bounding boxes for a given
loop nest, but we start with the tightest (smallest) bounding box among all
the possibilities.

The bounding box also plays an important role in deciding the leaf tile size.
We want the leaf tile size to have the exact value as the input tile size
parameter to the generated code. Therefore, we pad the size of bounding box
along each dimension to the minimum value of $b_i\times2^{k_i} \ge N_i$ where
$b_i$ input leaf size parameter, $N_i$ the size along the $i^{th}$ dimension
of bounding box and $k_i \ge 0$ is an integer. Now, at $k_i$th level of
recursion the orthant size along the $i$th dimension will be $b_i$. The
padding of the bounding box introduces iteration points outside of original
iteration space.  These empty points get optimized away by the optimizations
described in Section~\ref{sec:codegen:opt}

\section{Empirical Study}
\label{sec:experiments}
In this section, we empirically study different tiling strategies. The goal of
the experiments is to characterize SLT and COT, and to show the influence of the
factors discussed in Section~\ref{sec:motivations}. 



\subsection{Experimentation Setup}
For our experiments, we use two CPU architectures: Haswell and Broadwell. The
configuration of these machines are shown in Table~\ref{tab:machine}.  The
benchmarks are compiled with ICC 16.0.2 using the following flags: \texttt{-O3
-xHost -ipo}.  We use \texttt{perf} tool to count off-chip memory accesses
(OCA), which is the total number of LLC misses. The tool counts the number of
OCA for the entire program that consists of kernel, timing functions and reading
program inputs. The number of OCA is dominant in the kernel, therefore we can
safely use \texttt{perf} to measure OCA for the kernels.

\begin{table}
\caption{Machine Configuration
\label{tab:machine}}
\centering
\begin{tabular}{|l|r|r|r|}
\hline\hline
 Architecture Parameters &Haswell & Broadwell \\ \hline\hline
 Processor &  E3-1231 v3 & E5-1650v4 \\ \hline
 Base Frequency & 3.4 GHz & 3.6 GHz \\ \hline
 Turbo Boost Frequency & 3.8 GHz & 4 GHz \\ \hline
 Number of Cores &  4 & 6 \\ \hline
 RAM & 32 GB & 16 GB \\ \hline
 L3 Cache &  8 MB 16-way & 15 MB 20-way\\ \hline
 L2 Cache &  \multicolumn{2}{c|}{256 KB per core 8-way} \\ \hline
 L1 Cache &  \multicolumn{2}{c|}{32 KB per core 8-way} \\ \hline
 Max Mem.Bandwidth & 25.6 GB/s & 76.8GB/s \\ \hline
 Compiler & \multicolumn{2}{c|}{icc 16.0.2} \\ \hline
\end{tabular}
\end{table}

\subsubsection{Code Generators}
We experiment with multiple state-of-the-art code generators.  We use
D-Tiling~\cite{sanjay-lcpc2009,sanjay-kim-dtilingTR-2010} implemented in
AlphaZ~\cite{yuki2013alphaz} to generate SLT code with parametric tile sizes.
We compare the performance of parametric SLT and PCOT, which are main subject
of our study. In addition, we compare speed of PCOT with
Pluto~\cite{uday-pldi08} for fixed size SLT code, Pochoir~\cite{Tang2011} for
cache oblivious Jacobi-style stencils, and Autogen~\cite{autogen-ppopp16} for
cache oblivious dynamic programming.

\subsubsection{Benchmark Suite}
We use kernels from stencils, linear algebra, and dynamic programming for our
experiments.  Specifically, we include FDTD-2D, GaussSeidel-2D (GS-2D),
Heat-2D and Heat-3D for stencils; Cholesky decomposition and LU decomposition
(LUD) for linear algebra, and Optimal String Parenthesizing (OSP) from the
dynamic programming.  
These kernels are selected to cover the types of kernels handled by existing
tools, as well as additional polyhedral kernels that can now be supported with
the PCOT generator.  All the benchmarks are from PolyBench/C
\footnote{Available online at
\url{https://sourceforge.net/projects/polybench/}} version 4.2 except for OSP.
We include OSP, since it was used to evaluate Autogen~\cite{autogen-ppopp16}
and we obtained OSP code from authors of Autogen.
All benchmarks are scheduled using the Pluto algorithm, and use the same
memory allocation as PolyBench kernels.  

Pluto algorithm is unable to find tiling hyperplanes for all three dimensions
for OSP. However, there is a well-known transformation that reorders the
summation described by Guibas et al.~\cite{guibas-kung-thompson} to transform
the dependences to be tileable.  We apply this transformation as a
pre-processing to tile all three dimensions of OSP. Autogen is able to tile all
three dimensions automatically,
making this necessary to make a fair comparison of the tile execution order.

\subsubsection{Problem Size Selection}
We select the problem size of each benchmark such that the memory footprint of
a wavefront of tiles, i.e., the set of tiles that may be executed in parallel
with single-level tiling code, does not fit in LLC. Therefore, during the
execution there will be off-chip memory accesses. We do not use problem sizes
that are  powers of 2 since they exacerbate the occurrences of conflict misses
unless padded.  Table~\ref{tab:bench} shows the chosen problem sizes for all
the benchmarks.

\subsubsection{Consistent Optimization Level Across Codegen Techniques} 
We want to ensure that the performance of each tile is similar across all
techniques to focus on the difference due to tile execution order.  Autogen
and Pochoir apply some low-level optimizations not supported by the PCOT
generator and Pluto. These optimizations were applied manually to SLT, PCOT,
and Pluto generated code.
%
These optimizations are:
\begin{itemize}
\item Copy optimization: copying transpose of a column-major input matrix
inside a base function to a local array, so that it can be accessed in unit
stride, and 
\item Write optimization: if there is an accumulation in the innermost loop,
first perform the accumulation on a local variable and then update the array
at the end.
\item Indexing Simplification: array accesses are simplified by use of pointer arithmetic
instead of accesses to multi-dimensional arrays.
 \end{itemize}
%
%
In addition we used the restrict keyword and additional compiler options to
ensure that parallel innermost loops are vectorized by ICC.

\begin{table}
\caption{Benchmarks, problem sizes and achieved performance in GFLOPs per
second for the sequential single-level tiling code on both architectures.
\label{tab:bench}}
\centering
\begin{tabular}{|l|c|S|S|}
\hline\hline
 \multirow{2}{*}{Benchmark} & \multirow{2}{*}{Problem size} &
\multicolumn{2}{c|}{{SLT-seq GFLOPS}} \\  \cline{3-4}
  &  & {Haswell} & {Broadwell} \\ \hline\hline
 Cholesky & $4000^2$ & 12.3 & 11.7 \\ \hline
 LUD & $3000^2$ & 4.1 & 4.1 \\ \hline
 FDTD-2D & $500\times 2000^2$ & 6.4 & 5.6 \\ \hline
 GS-2D & $500\times 2000^2$ & 1.9 & 2.7 \\ \hline
 Heat-2D & $500\times 3000^2$ & 20.5 & 20.4\\ \hline
 Heat-3D & $50\times 350^3$ & 16.5 & 17.6\\ \hline
 OSP\footnotemark & $4000^2$ & 21.0 & 22.8 \\ \hline

\end{tabular}
\end{table}
\footnotetext{OSP performance numbers are in GUpdateS}

\subsubsection{Tile Size Exploration} 

%
In our study, we explore the difference between SLT and COT. An
important aspect of this study is the influence of tile sizes, which have
large impact on both approaches.  We are not interested in the ``best'' tile
size in this paper, but in the characterization of the effect of tile sizes.
Therefore, our main focus is in the statistical property over a same set of
tile sizes for both tiling schemes. 
%

We explore a set of tile sizes ranges from 10 to 50 in steps of 10, 100 to 500
in steps of 100 in all the dimensions, and 1000 to problem-size in steps of
1000 along innermost dimension of a tile. For Heat-3D, tile sizes 10, 30, 50,
100 and 300 were explored along all the dimensions. These ranges of tile sizes
give a relatively coarse-grained sampling of possible tile sizes that fits in
different levels of cache. Exhaustively searching the entire space of legal
tile sizes is impractical due to time limitations.  For each tile size, we use
the mean over three runs, which had low variance across runs.

We do not explore tile sizes on both Pluto and Pochoir. It is impossible to
automatically explore tile sizes for Pluto since it performs fixed sized
tiling, and we had to manually modify the output to make the performance of
tiles consistent with others. The tile sizes in Pochoir are hardwired into the
code generator, and we could not find an easy way to explore the tile sizes.


\begin{figure*}[t!]
	\centering
	\begin{subfigure}{1\textwidth}
			\centering
			\includegraphics[width=1\textwidth]{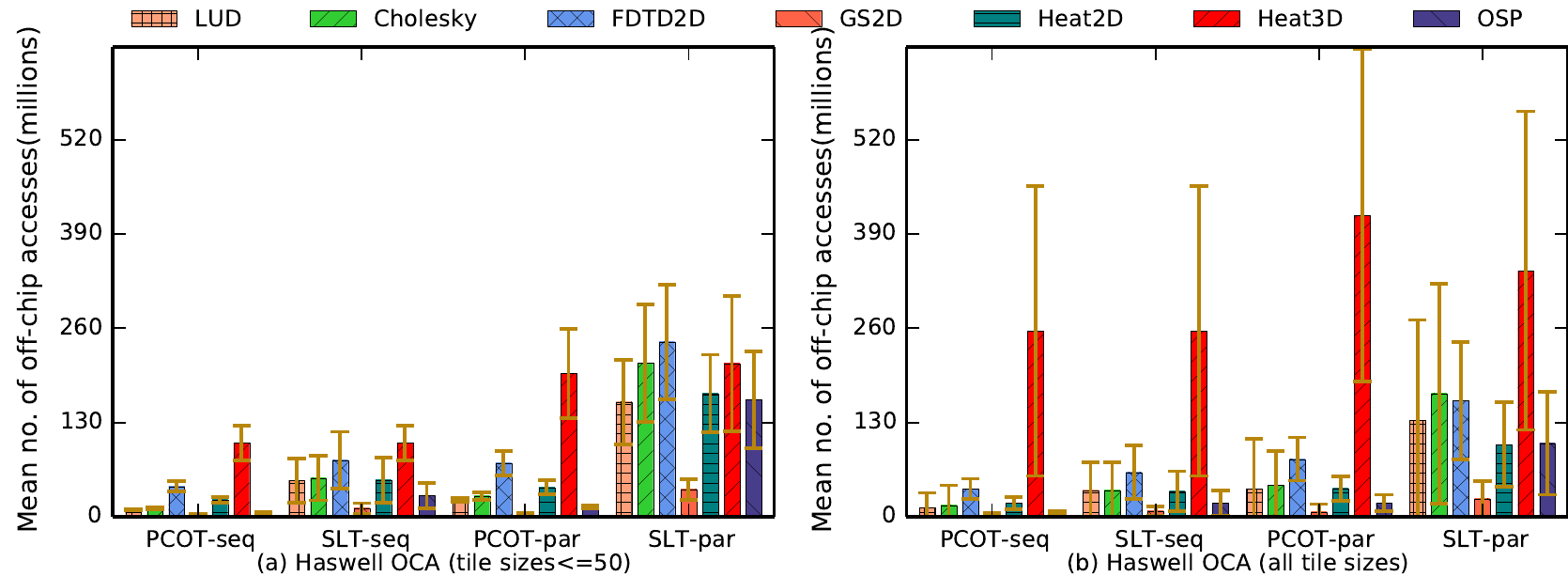}
	\end{subfigure}%
	 
	\begin{subfigure}{1\textwidth}
			\centering
			\includegraphics[width=0.988\textwidth]{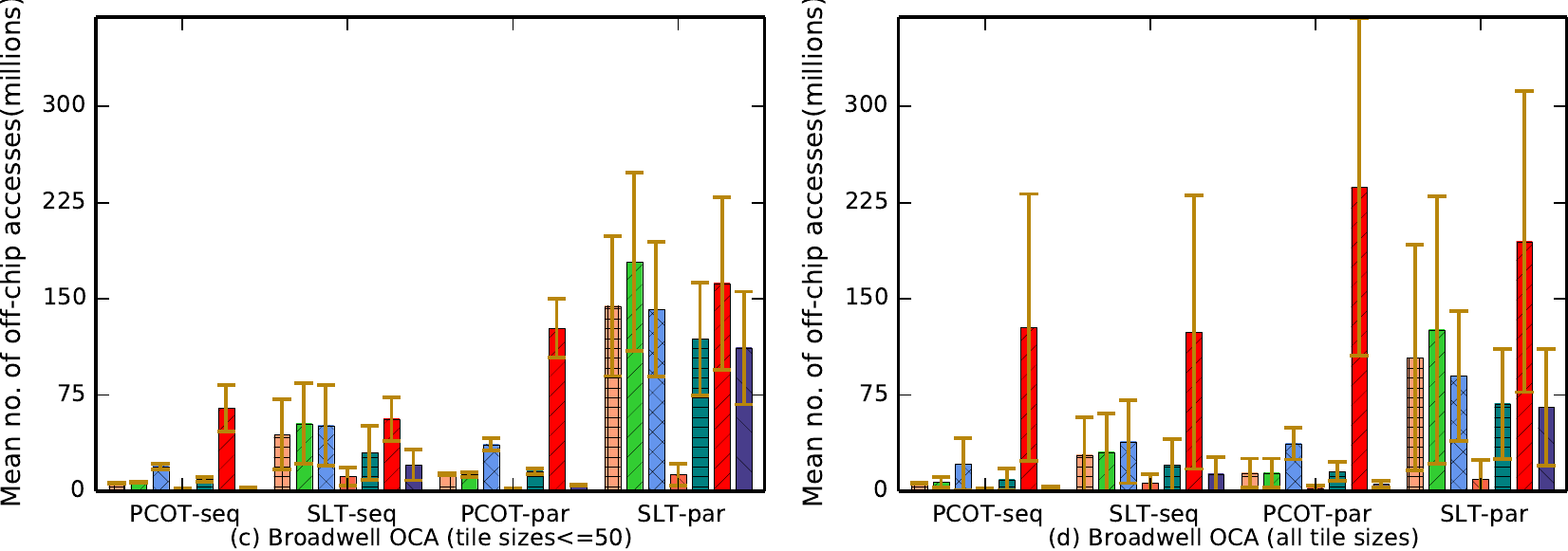}
	\end{subfigure}
	\caption{Mean and standard deviation of number of OCA across the tile sizes
that fits in LLC for both Haswell and Broadwell. The figures in the left (a,c)
show the OCA behavior when restricted to tile sizes within 50 in all
dimensions and the figures in the right (b,d) show the behavior when all the
tile sizes are used. We have to focus on both relative standard deviation and
mean number of OCA. When the tile sizes are restricted to 50, we see the
expected behavior where the variability is much lower with COT. This confirms
the benefits of the recursive structure of COT as it automatically takes
advantage of multiple levels of memory hierarchy.  When the tile sizes are not
restricted, the number of OCA remains similar for COT. The mean number of OCA
for SLT decreases compared to smaller tiles since the larger tiles fits in LLC
reducing the number of OCA.  The behavior of OCA on both cpu platforms is
similar except that Haswell has relatively more OCA since its LLC size is
smaller compared to Broadwell.
} 
	\label{fig:oca_var}
\end{figure*}

%

\begin{figure*}[tb]
  \centering
    \includegraphics[width=0.98\textwidth]{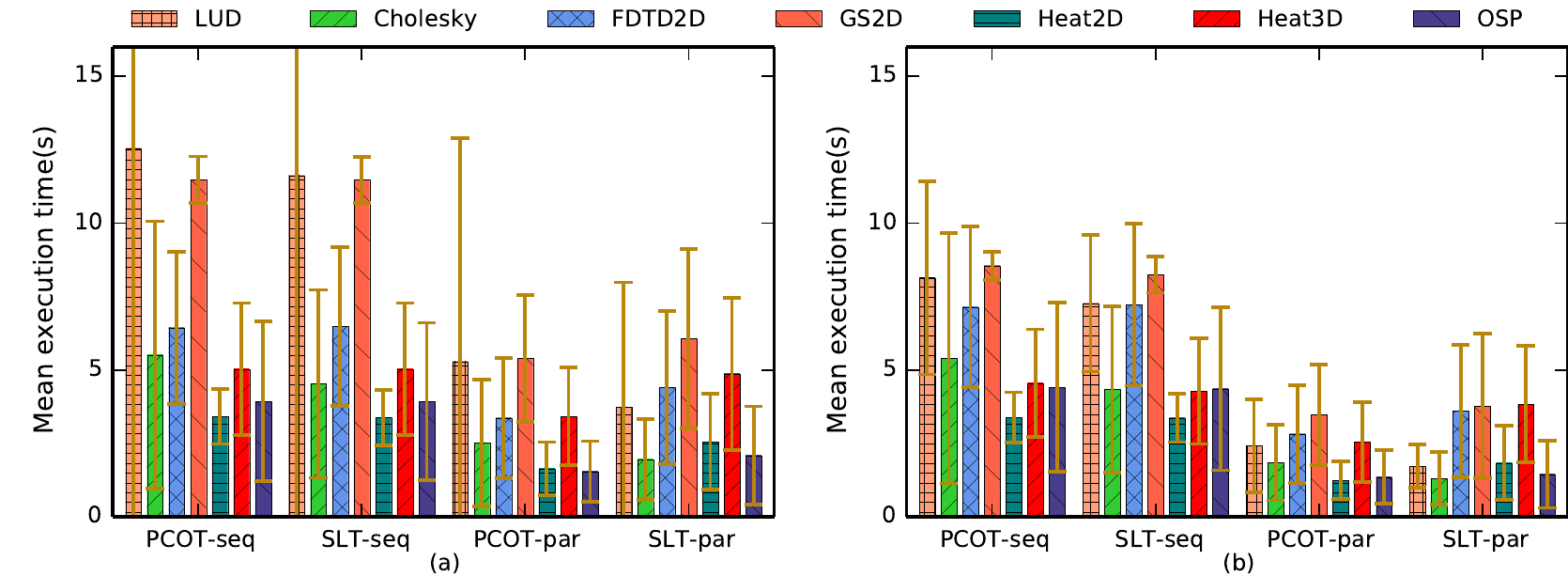}
\caption{Mean and standard deviation of execution time on (a) Haswell and (b)
Broadwell across all explored tile sizes. Variance is high for both COT and
SLT on both machines. Therefore, tile size exploration/tuning is needed for
speed.}
	\label{fig:ex_avg_all_var}
\end{figure*}

\subsection{Characterization of SLT and COT}
Figures~\ref{fig:oca_var}, and \ref{fig:ex_avg_all_var} summarize the result
of our study in the two platforms.

We first discuss the variability of OCA with respect to tile size. COT has
much smaller number of OCA and variability compared to SLT when we restrict to
smaller tile sizes (Figures~\ref{fig:oca_var}a and~\ref{fig:oca_var}c). This
confirms the benefits of COT as it automatically takes advantage of the
multiple levels of memory hierarchy, owing to the recursive structure of COT. 

Heat-3D does not exhibit the expected behavior since COT introduces more
conflict misses for higher dimensional data as explained below.
%
The theoretical results of COT were derived with the assumption of fully
associative caches where a cache line can be stored at any location in the
cache. Therefore, there will not be any occurrences of conflict misses but
capacity misses. In reality, caches are n-way (n=16 and 20 for the cpus used
in our experiments) associative where there are n candidate locations in the
cache to store a cache line. This leads to conflict misses. 
Let us consider sequential execution of a d-dimensional stencil where all d+1
(including time dimension) dimensions are tiled.  A tile uses d of the faces
as inputs and generates d faces of outputs.  Let us assume that the tile sizes
are such that data correspond to all d output faces fit in LLC. 
SLT executes tiles in lexicographic order.  Therefore, lexicographically next
tile will read one of the input faces through LLC. But all other d-1 input
faces were produced long time ago, therefore data is no longer in LLC due to
potential capacity misses. Hence, conflict misses can potentially occur when
the input face is read through LLC and when accessing data while computing
values within the tile. COT visits the neighbouring tiles first. Therefore, a
tile reads more than one (ideally d) input faces through LLC. Hence, conflict
misses can occur when all the input faces are read and when compute values
within the tile. Since COT reads more input faces through LLC, there is a
higher chance of occurring conflict misses compared to SLT.  When the number
of dimensions are increased, the dimensions of the  memory footprint of a tile
increase as well as the number of input faces COT read through LLC. The higher
number of dimensions, the larger the distance between memory footprints of
neighbouring tiles. Therefore, the chance of occurring conflict misses is
further increased. Hence for Heat-3D, the number of conflict misses for COT is
as high as the number of capacity misses for SLT. This behaviour diminish the
benefits of COT for higher dimensional stencils.


When data points correspond to all the tile sizes are considered
(Figures~\ref{fig:oca_var}b and ~\ref{fig:oca_var}d), the absolute number of
OCA remains similar for COT since all the tile sizes fits in LLC. For SLT, the
mean of OCA is slightly decreased compared to smaller tile sizes since the
larger tiles fit in LLC reducing the number of LLC misses. Heat3D remains an
outlier due to the conflict misses as described in previous paragraph. The
behavior of OCA we discussed so far similar for both sequential and parallel
codes except that parallel codes have more OCA due to the increased number of
conflict misses.



The variability of execution time is high for both COT and SLT on both
platforms. It is related to the recursion overhead and restrictive tile sizes
due to recursive split of COT as described in Section~\ref{sec:motivations}.
This shows that speed of COT is sensitive to tile sizes, therefore, both COT
and SLT need tile size exploration/tuning for speed. For sequential runs,
large tile sizes make more effective use of prefetcher with increased
bandwidth requirement as described under \emph{hardware prefetching} in
Section~\ref{sec:motivations}. We verified this behavior of prefetcher by
counting the L2 prefetcher activities. We noticed that for large tile sizes
the number of prefetcher activities are 3 order of magnitude higher compared
to small tile sizes.  But for parallel runs, the bandwidth limit is reached
faster, therefore, benefits of prefetching diminish.

%
%
%
%
%

\begin{figure*}[tb]
  \centering
    \includegraphics[width=0.98\textwidth]{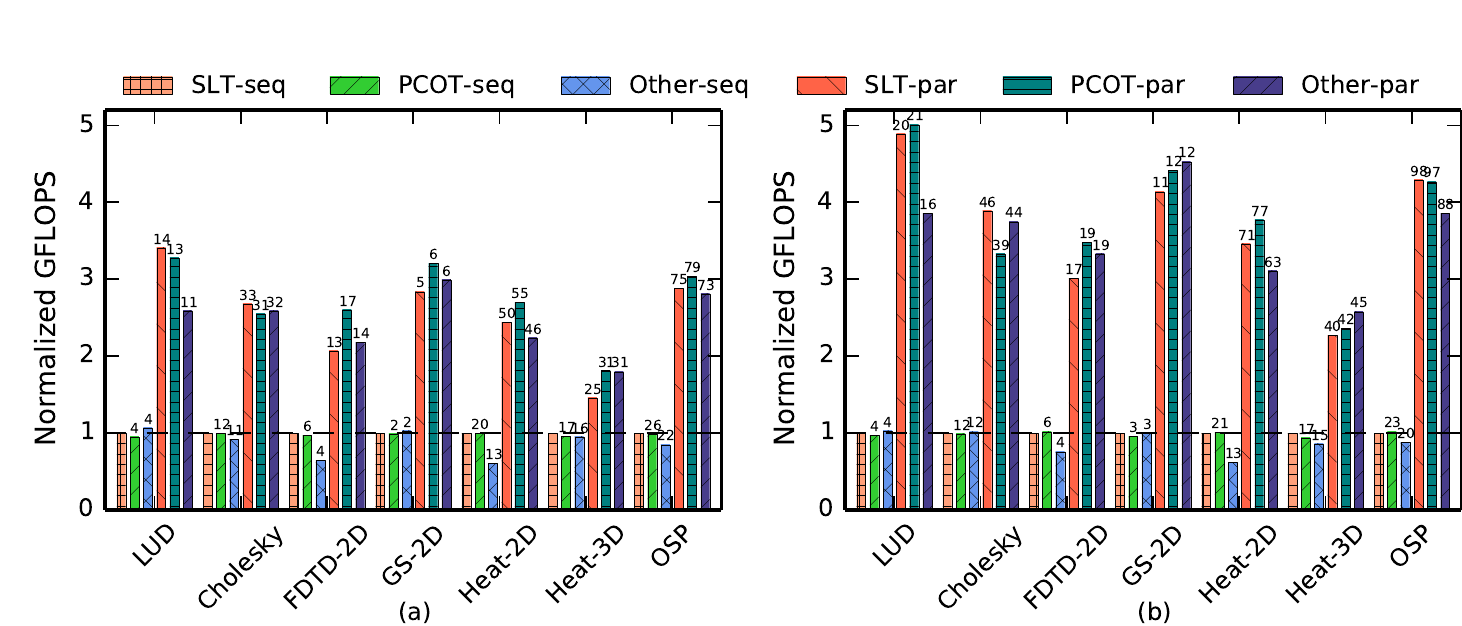}
\caption{Performance of SLT and PCOT on (a)
Haswell and (b) Broadwell, normalized to sequential parametric SLT
\texttt{SLT-seq} (higher the better). The absolute performance numbers are
shown on top of each bar in GFLOPS. Both the sequential and the parallel
performance of SLT and PCOT closely matches the performance of other code
generators. \texttt{Other} represents Pluto for the first 4 benchmarks, Pochoir
for Heat-2D and 3D, and Autogen for OSP.}
	\label{fig:gflops}
\end{figure*}

\subsection{Comparison of Speed Against State of the Art}
In the following we compare the performance of SLT and PCOT with
state-of-the-art code generators.  The performance is normalized to the GFLOPS
(giga flops per second) achieved by sequential parametric single-level tiling
(\texttt{SLT-seq}) code.  We use the best tile size found in the tile size
exploration described above. For Pluto, we used the best tile size for SLT.
The absolute performance numbers are shown on top each bar in Figure~\ref{fig:gflops}

The parallel variants are executed with the number of threads equal to the
number of physical cores (i.e., hyper-threading is not considered) on each
machine.  Figure~\ref{fig:gflops} shows the normalized performance (GFLOPS) on
both Haswell and Broadwell. The speed of SLT and PCOT closely matches the
speed of state-of-the-art.

PCOT out performs Pochoir on Heat-2D. For this benchmark, Pochoir does not
vectorize the iterations at the boundaries. Pochoir scales better (20\%) with
6 threads for Heat-3D due to the concurrent start of tiles. 
For OSP, PCOT is 10-15\% faster than Autogen. This difference is attributed to
the tile shapes. The Autogen generated code is restricted to cubic tile sizes
where as the PCOT generated code have more degrees of freedom, and we found
that using large tile sizes in the innermost dimension is better for
performance in many cases. When we try cubic tile sizes in PCOT, the speed is
similar to Autogen.
Performance of SLT, Pluto and PCOT are similar with each other.

Despite the significant reduction in off-chip memory accesses of benchmarks in
PCOT, the absolute speeds of codes from PCOT and other techniques remain
similar. For compute-bound kernels (i.e., after applying SLT), the latency of
OCA are already hidden. Therefore, further reducing OCA does not translate
into savings in execution time. 

Finally we compare the speed (GFLOPS) of the benchmarks to the peak
performance per benchmark. The compiler does clever optimizations for both
Haswell and Broadwell architectures. For example, 5-point Heat-2D stencil has
4 muls, 2 subs and 4 adds and the throughput of vector (4 doubles wide)
instructions is 2, 1 and 1 per cycle respectively \footnote{Data available at
http://software.intel.com/sites/landingpage/IntrinsicsGuide/}. 
The compiler replace some of the operations with fused multiply-add/sub (FMA)
operations where the throughput is 2 per cycle. The resulting assembly code
has only 2 subs, and 4 FMAs.  Hence, it takes only 4 cycles to execute 6
vector instructions which corresponds to 40 scalar operations. Therefore,
achievable peak (ceiling) is 36 GFLOPS on Broadwell. The achieved performance
is 20.5 GFLOPS which is 57\% of the peak. If we account for the cycles for the
non-arithmetic operations in the loop body of the assembly code, then the
achievable ceiling will be further lowered making the achieved performance
much closer to the peak.

\section{Related Work}
\label{sec:related}

We put our work into context and discuss other approaches to cache-oblivious methods. We also discuss previous works on tiling and code generation. 

\subsection{Empirical Study of Cache-Oblivious Methods}
The most closely related work is by Yotov et al.~\cite{yotov2007experimental} that explored answers to the question ``what is the cost of cache obliviousness?'' Their study is primarily on matrix multiplication and is only about sequential execution.  Our work extends the study to a broader range of programs and to parallel executions. We also provide variability of both speed and cache behavior with respect to tile size.

Some prior work on code generators for various types of tiling have empirically compared iteration space tiling and cache-oblivious methods~\cite{Bandishti12, lifflander2017cache, autogen-ppopp16, strzodka2010cache, zou2015rajopadhye}. The objective of these experiments are slightly different from ours; the main target of evaluation is the code generation tools. Our interest is in understanding the difference due to the tile execution order. The reported performance difference in these experiments mostly come from differences in other aspects that influence the performance (e.g., copy optimization and other low-level optimizations), which we carefully checked that codes from all tools have similar behavior in our work. This explains the apparent inconsistency with earlier results.

\subsection{Cache Oblivious Code Generation}
Prokop~\cite{prokop-thesis99, frigo-etal-focs99} introduced the cache-oblivious algorithms. Later, Frigo and Strumpen proposed the serial~\cite{frigo-strumpen-ics05} and the parallel~\cite{frigo2006cache} divide and conquer implementations of stencil programs.  Strzodka et al. ~\cite{strzodka2010cache} proposed cache oblivious parallelograms method (CORALS) which is similar to the time-skewed wavefront parallelization of iterative stencil computations. CORALS takes advantage of the regular dependence patterns of stencil computations and applies oblique cuts in both space and time dimensions simultaneously. They use a load-balancing scheme to evenly distribute all the threads amongst the tiles.  Their technique benefits from data locality, parallelism and vectorization simultaneously.  

Pochoir~\cite{Tang2011} is a domain specific compiler for stencil programs. It generates a divide and conquer implementation of the program based on trapezoidal decompositions using hyperspace cuts.  Pochoir generates efficient code, but the code generated for the hyper-trapezoidal tile shape is very complex and that the base case sizes are fixed at compile time.  Pochoir, also, cannot handle dependences along the time dimension.  The dependence should be always from a previous time step. Therefore, Gauss-Seidel-like dependences cannot be handled. Pochoir can generate the divide and conquer code for periodic stencils.  Periodic stencils can be handled by our COT code
generator after the smashing transformation ~\cite{sanjay-lcpc08,
uday-pact2014}.  Pochoir is capable of applying oblique cuts whereas we make canonic cuts.

Serial and parallel implementations of recursive divide and conquer algorithms with optimal cache complexity have been developed and evaluated for a specific dynamic programming algorithm such as Longest Common Subsequence ~\cite{rezaul-thesis, Chowdhury-2006-CDP}, global pairwise sequence alignment problem in bioinformatics ~\cite{Chowdhury-2010soda}, Gaussian Elimination Paradigm ~\cite{Chowdhury2010GEP}, etc. Tithi et al.~\cite{tithi-ipdps2015} described a way to obtain divide and conquer algorithms manually for a class of dynamic programming problems. The base cases of these programs are similar to matrix multiplication computations. The code is hand optimized with the non-trivial Z-Morton matrix conversions.  Unlike the work of ~\cite{tithi-ipdps2015}, Autogen does not use Z-Morton matrices. The derived cache-oblivious implementation is parameterized by a single base value for all the dimensions which limits the tile shape to a hypercube.  Autogen is mostly applicable to dynamic programming problems where the fractal property is often available.

Tang et al.~\cite{Tang-COW} proposed the Cache-Oblivious-Wavefront (COW) technique, that improves the parallelism of the cache-oblivious algorithms, which also preserves locality. They use classic divide and conquer strategy to partition the iteration space and schedule the execution of tasks across different levels of the recursion as soon as the data dependency constraints are satisfied. The execution pattern of the COW algorithm is conceptually similar to classic wavefront schedules. However, the implementation is done by hand and their technique can handle only dynamic programming algorithms.

Autogen~\cite{autogen-ppopp16} executes an iterative implementation of the input program using very small problem sizes to determine the access patterns (dependences).  It then determines a recursive decomposition (called fractal property) that can be used to derive a divide and conquer implementation. The structure of the recursive function is automatically generated but the body of the base case is hand implemented.  

Bellmania~\cite{itzhaky-oopsla2016} is an interactive system, based on solver-aided tactics, that re-writes the rules and generates provably correct divide and conquer implementations of dynamic programs.  The dependences are statically analyzed, however, the derivation of the recursive algorithm is not automatic.

The COT code generator presented in this paper takes the base case tile sizes as input, therefore, the tile shape can be hyper-rectangular; and supports more general class of programs, i.e. all polyhedral programs.

\subsection{Tiled Code Generation}
Tiling~\cite{Wol87,irigoin-popl88} is a classic iteration space partitioning technique which combines a set of points into tiles, where each tile can be executed atomically.  Tiling comes in handy for exploiting data locality~\cite{Wolf91tiling, kamil2010, liu2009}, minimizing communication ~\cite{Andonov2001,xue-jpdc97} and maximizing parallelism ~\citep{Bandishti12}.  

Time skewing ~\cite{wonnacott-time-skewing-tr99a} enables time tiling which increases the temporal reuse, especially in stencils.  Bondhugula et al. ~\cite{uday-pldi08} developed PLuTo, a system for tiling imperfectly nested affine loops with fixed sized tiles.  The PLuTo algorithm obtains tiling hyperplanes that minimizes communication across fixed size hyper-parallelopiped tiles. Bandishti et al. ~\cite{Bandishti12} proposed a diamond tiling technique that enables the concurrent start of tiles and eliminates the pipeline fill and flush cost of classic wavefront tiling. Many other tiling techniques ~\cite{Hartono-DynTile, grosser2014hybrid, TJin-Hybrid-Tech, Krishnamoorthy07, Henretty2013SIMDStencil, Holewinski2012Overtile} also propose ways to eliminate pipeline start-up cost in a similar fashion.  

However, in all of the above, the generated code has fixed sized tiles and does not guarantee optimal usage of caches.  Parametric tile sizes enable efficient auto-tuning and performance portability. Techniques ~\cite{Hartono2009prime, baskaran-etal-cgo10, Kim2010, lakshmi-thesis, sanjay-kim-dtilingTR-2010, darte2014parametric}  have been thus developed to allow parametric tiling. More studies on parametric tiling such as D-Tiling ~\cite{sanjay-lcpc2009, sanjay-kim-dtilingTR-2010}, P-Tiling~\cite{baskaran-etal-cgo10} and Mono-parametric tiling~\cite{iooss:hal2015} are conducted where the latter is a polyhedral transformation.  

Our COT code generator produces code with parametric base tile sizes and the point loops can be generated using any of the aforementioned parametric tiled code generators.

\section{Conclusion and Future Work}
\label{sec:conclusion}

In this paper, we have presented an empirical study of two approaches for
tiling: SLT and COT. We have developed cache oblivious code generator that
support any polyhedral computation to widen the scope of study to polyhedral
programs.  The takeaway is that COT does not save you from tuning effort for
speed, but it gives you reduced off-chip memory accesses without tuning.
However, when the speed is the primary objective, situations where good
utilization of hardware prefetcher is essential negate the benefit of COT.

It is interesting to note that the access pattern of Jacobi-style stencils,
which has been one of the main beneficiary of COT, is prefetcher-friendly, and
are not fully benefiting from COT in our study. Our code generator now offers
the option to use COT instead of other execution strategies implemented in the
polyhedral tools for a much wider class of programs than stencils.  

Since the speed of PCOT and SLT are similar, the savings off-chip memory
accesses may result in saving in energy. It would be interesting to quantify
the energy savings as a future work.  This work only investigated Single-Level
Tiling where multi-level tiling is a known alternative to improve data
locality at multiple levels of the memory hierarchy.  Further study that also
explore the use of multi-level tiling is a natural future work.


\bibliographystyle{ACM-Reference-Format}
\bibliography{TACO2017}

\appendix
\section{Schedule Independent Memory Allocation}
\label{sec:sima}

We also address a memory-based limitation of polyhedral compilation
tools.  It is well known that in any parallelization (of any program), it is
essential to respect (only) the \emph{true} or flow dependences.  Other
(memory-based) dependences can be ignored if one can re-allocate memory.  In
practice, this is limited by the fact that the associated memory expansion may
be prohibitively expensive, and there has been work on mitigating this
expansion~\cite{vasilache-impact12, lefebvre-feautrier-pc98, sanjay-europar96,
  sanjay-toplas00}.  We propose a novel yet simple \emph{schedule-independent}
memory allocation strategy.  Our work also generalizes polyhedral compilation
by enabling polyhedral tools to use alternate, \emph{hybrid} schedules
consisting of affine loops for certain parts of the iteration space and
cache-oblivious divide-and-conquer schedules for others.

\subsection{Background}

In this section, we introduce the necessary background of our work. We first
give a brief description of the polyhedral representation of programs, and the
general flow of a polyhedral compiler.  Then, we discuss the legality of
tiling, which is related to the input of our code generator.

\begin{figure*}[tb]
  \centering 
  \includegraphics[scale=0.6]{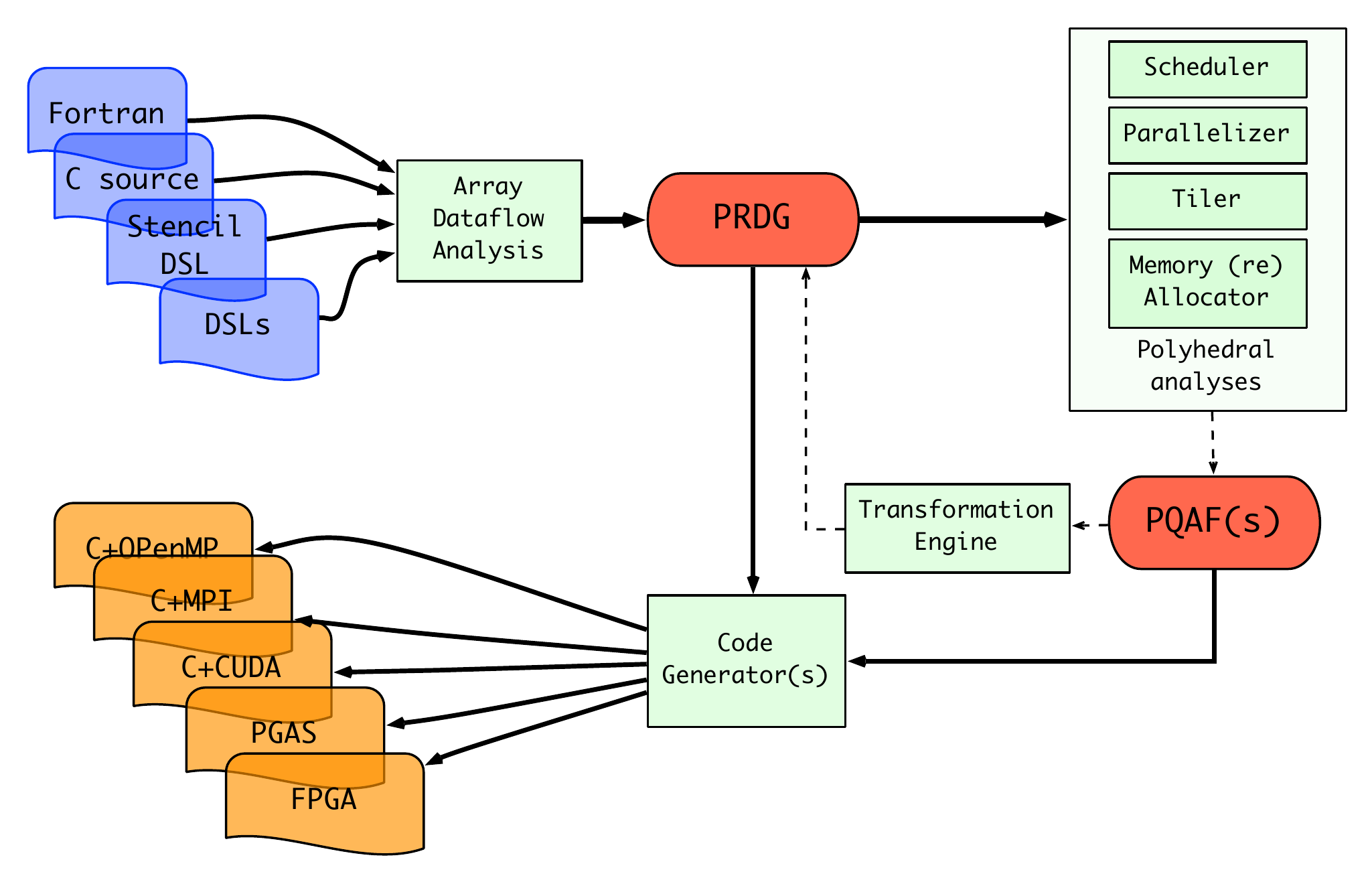}
  \caption{\small{Polyhedral Compilation: the Polyhedral Reduced Dependence
      (hyper) Graph (PRDG) serves as the intermediate representation.
      Piecewise Quasi-Affine Functions (PQAFs) describe transformations.}}
  \label{fig:compiler}
\end{figure*}

\subsubsection{Polyhedral Compilation and Representation}

Figure~\ref{fig:compiler} shows the flow of polyhedral compilation.  First,
dependence analysis of an input program (or a ``polyhedral section'' thereof)
produces an intermediate representation (IR) in the form of~\cite{DRV-sched00}
a \emph{Polyhedral Reduced Dependence (hyper) Graph} (PRDG).  Various analyses
are performed on the PRDG to choose a number of mappings in the form of
\emph{Piecewise Quasi-Affine Functions} (PQAFs) that specify the schedule as a
set of \emph{multi-dimensional} vectors.  The PQAFs come with annotations to
indicate whether each dimension is sequential or parallel, and also whether it
is part of a \emph{tilable band}, i.e., whether tiling this band of dimensions
is legal.  The transformations may be applied to the PRDG iteratively, and
(eventually) the PRDG and QLAF are provided to a code-generator that produces
code for various targets.


One of the strengths of the polyhedral model is that a parametric program may
be concisely represented with a PRDG with finite number of nodes (statements)
and edges (dependences).  The potentially unbounded sets of instances of a
statement are represented in abstract forms of integer sets, called
\emph{domains}, and dependences between them as affine functions (or
relations, which are viewed as a set-valued function) over these statement
domains.  Indeed, every edge, $e$ from node $v$ to $w$, in the PRDG is
annotated with two objects: (i) a domain, $D_e$ specifying the (subset of) the
domain, $D_v$ of its source node, where the dependence occurs, and (ii) the
affine function, $f$, such that for any point $z\in D_e$, the (set of)
point(s) in $D_w$ on which it depends is given by $f(z)$.  $D_e$ is called the
context of the edge, and $f$ is its dependence function.  We also use the
notation $f(D_e)$ to denote the set valued image of $D_e$ by $f$.

An affine function $\mathbb{Z}^n \rightarrow \mathbb{Z}^m$ may be expressed as
$f(x) = A\vec{x} + \vec{b}$, where $\vec{x}$, function domain, is an integer
vector of size $n$; $A$, linear part, is an $n\times m$ matrix; and $\vec{b}$,
constant part, is an integer vector of size $m$.  A dependence is said to be
uniform if the dependence function is only a constant offset, i.e., when the
linear part $A$ is the identity.


\subsubsection{Legality of Tiling}

Tiling is a well-known loop transformation for partitioning computations into
smaller, atomic (all inputs to a tile can be computed before its execution),
units called tiles~\cite{irigoin-popl88, Wolf91tiling}.  The natural legality
condition is that the dependences across tiles do not create a cycle.  In
compilers, this condition is typically expressed as fully permutability (i.e.,
dependences are non-negative direction vectors), which is a sufficient
condition.  Our transformation for cache oblivious tiling takes as inputs a
loop nest that is fully permutable.  For polyhedral programs, scheduling
techniques to expose such loop nests are available~\cite{uday-pldi08}.

\subsection{Memory Allocation}

\begin{figure}[tb]
  \centering 
{\small\begin{lstlisting}
for (i = 0; i < N; i++){
  S0:  X[0,i] = A[i]; } // Initialize
for (t = 1; t <= 2*N; t++){ //Note: ub is even
  for (i = 1; i < N-1; i++){
    S1: X[t%2][i] = f(X[(t-1)%2][i-1],
             X[(t-1)%2][i], X[(t-1)%2][i+1],
             X[(t-1)%2][0]);
    }
  S2: X[t%2][0] = g(X[(t-1)%2][0]); 
  S3: X[t%2][N-1] = g(X[(t-1)%2][N-1]);
}
for (i = 0; i < N; i++){
  S4: Aout[i] = X[0,i]; } // Output copy
\end{lstlisting}
}
\caption{\small{Neither Pochoir nor Autogen can handle the computation
    performed by this simple loop.  Moreover, it has a memory based dependence
    that prevents polyhedral compilers like Pluto from tiling both dimensions.
    However, the true dependences of the program admit a tilable schedule, but
    at the potentical cost of $O(N^2)$ memory.  Our scheme reduces this to
    $O(N)$}}
\label{fig:motiv}
\end{figure}

In this section, we first describe how memory based dependences prevent
tiling, using our motivating example (Fig.~\ref{fig:motiv}), and show that
simply ignoring these (false) dependences would lead to memory explosion.
After formulating our problem, we next propose a simple, schedule independent
memory allocation scheme that resolves it.

Consider the statement $\mathrm{S1}$, and note that its domain, $D_1$ is the
polyhedral set, $\{t,i~|~ 1\leq t\leq 2N \wedge 1\leq i\leq N-1 \}$.
$\mathrm{S1}$ has four true dependences (for points sufficiently far from the
boundaries), three of which are $\mathrm{S1}[t-1, i-1]$, $\mathrm{S1}[t-1, i]$
and $\mathrm{S1}[t-1, i+1]$, the typical, 1D-Jacobi stencil dependences, and
the fourth one is $\mathrm{S2}[t-1, 0]$, which is a truly affine dependence on
the most recent writer to the memory location $\mathtt{X[(t-1)\%2,0]}$ when
the statement $\langle \mathrm{S1}, [t, i]\rangle$ is being executed.  All
these dependences are captured as edges with affine \emph{functions} in the
PRDG.  In addition, there is a memory based dependence, that we must also
respect.  Consider statement $\mathrm{S2}$, whose domain, $D_2 = \{t~|~ 1\leq
t\leq 2N\}$ is just one dimensional.  The $t$-th instance of S2 \emph{(over)
  writes} $\mathtt{X[t\%2, 0]}$, therefore all computations that read the
previous value must be executed before it.  In this sense, $\mathrm{S2}[t]$
``depends on'' the set $\mathrm{S1}[t,i]$, for all $1\leq i\leq N-1$.  This
dependence (which is a \emph{relation} rather than a function) is captured by
another a special edge in the PRDG.

The only schedule that respects all these dependences is the family of lines
parallel to the $t$ axis (provided all iterations of S1 are done first).
Although this has maximal parallelism, it has very poor locality.  Note that
the Pluto scheduler does not seek maximal parallelism, but rather, to maximize
the \emph{number of linearly independent tiling hyperplanes}.  Unfortunately,
the $t=\mathrm{const}$ is the only legal tiling hyperplane for this set of
dependences, and the tilable band obtained by Pluto is only 1-dimensional.

What if we did not have the memory-based dependences, i.e., what if we ignored
the memory allocation of the original program, and stored each computed value
in a distinct memory location?  In this case, there would be no memory based
dependences, and we can indeed find another family of (actually, infinitely
many) legal tiling hyperplanes: say, the lines $i+t=\mathrm{const}$.  As a
result, if we use the mapping $(t,i) \mapsto (t,i+t)$ as our ``schedule,'' the
new loops in the transformed program would be fully permutable, and could be
legally tiled.

Thus, the problem we seek to solve is: \emph{how to avoid memory based
  dependences, but without the cost of memory expansion that it seems to
  imply}.

Memory allocation for polyhedral programs is a well studied problem, and there
are two main approaches.  One either does memory allocation after the schedule is
chosen~\cite{sanjay-europar96, degreef-memory97, lefebvre-feautrier-pc98,
  sanjay-toplas00, darte-lattice05, vasilache-impact12,
  bhaskaracharya-toplas16, bhaskaracharya-popl16} since it often leads to a
smaller memory footprint, or else uses a \emph{schedule independent} memory
allocation, based on the so called \emph{universal occupancy vectors} (UOV).
This problem is solved when the program has \emph{uniform dependences}, i.e.,
when each dependence can be described by a \emph{constant vector}, and for some
simple extensions of this~\cite{strout-etal-asplos98, sanjay-memory-2011}.

It is important to note that tiling actually modifies a schedule: the so
called, ``schedule dimensions'' become fully permutable loops, and indeed,
these loops \emph{are actually permuted} in the generated tiled code.  So,
when a \emph{tiling schedule} specified by a family of $d$ tiling hyperplanes
is finally implemented by the generated code, the actual time-stamps are not
really $d$-dimensional vectors, but rather $2d$-dimensional ones obtained as
some complicated function of these indices.  Furthermore, we will see when we
generate cache-oblivious tiled codes, these tilable loops will actually be
visited in the divide-and-conquer order of execution, as required by COT.  As
a result, finding a memory map that takes into account such a rather
complicated final schedule is a tricky problem.  We therefore seek and propose
schedule-independent memory allocations.

The intuition behind our solution is (deceptively) simple, and we first
illustrate it on our motivating example (Fig.~\ref{fig:motiv}).  Rather than
the so-called ``single assignment'' program for the entire iteration space of
the program (i.e., full memory expansion), could we find lower-dimensional
subsets, such that a single assignment memory for only these subsets is
sufficient?  A careful examination of the code reveals that the memory based
dependences arise due to statement S2, and its domain is only 1-dimensional.
So we store the results of this statement into an auxiliary array, \texttt{Y},
and modify the program so that the fourth dependence simply reads
\texttt{Y[t-1]}, rather than \texttt{X[(t-1)\%2,0]}.  For the variable,
\texttt{X}, we use the old $(t,i) \mapsto (t\%2,i)$ memory allocation that was
used in the original code.  This results in $4N$ memory, which is a polynomial
degree better than quadratic.  Of course, the challenge is how to discover
this automatically.

We now outline how this is done.  At a high level, our algorithm takes a PRDG
as input, applies some (piecewise) affine transformations to it, and outputs
the transformed PRDG together with a separate memory map for each node in the
transformed PRDG.  More specifically, it works as follows.

\begin{itemize}
\item \emph{Preprocessing.}  For each edge, $e$, in the PRDG, with context
  $D_e$, and function, $f$, we first identify whether $f$ is \emph{uniform in
    context} in the sense that, for all points, $z\in D_e$, the value of
  $z-f(z)$ is a constant vector, independent of $z$.

  For example, consider a dependence function, $(i,j) \mapsto (i-1,i-1)$
  which, maps any point $[i,j]$ in the plane to a point on the diagonal, and
  is clearly not uniform.  However, what if $D_e = \{i,j~|~i=j-1\}$?  With
  this contextual information, the dependence is actually uniform: $(i, j)
  \mapsto (i-1, j-2)$.

  All edges/dependences that are neither uniform to begin with, not uniform in
  context, are marked as \emph{truly affine}.
\item \emph{Affine Split.}  For every node, $v$, in the PRDG that has at least
  one truly affine edge $e$ incident on it, we create a new node, $v'$.  Its
  domain $D_{v'}$ is the union of $f(D_e)$ of all such incident edges.

  The edges in the PRDG are modified as follows.  All the truly affine edges
  that were incident on $v$ are now made incident on $v'$; and $v'$ has a
  single outgoing edge $e'$, annotated with $\langle D_{v'}, I \rangle$ (its
  dependence function is the identity map) and whose destination is $v$.

  It is easy to see that we have not changed the program semantics.  In
  effect, we have simply copied the value of every point in $D_v$ that was the
  target of any truly affine dependence over to a new variable $v'$, and
  ``diverted'' all the truly affine edges that used to be incident on $v$ over
  to $v'$.  Moreover, since the identity function is uniform by definition, all
  edges incident on $v$ are now either uniform, or uniform in context.
\item We now use existing UOV based methods~\cite{strout-etal-asplos98,
    sanjay-memory-2011} to choose a schedule-independent memory allocation for
  all the original nodes in the PRDG, and a \emph{single-assignment} memory
  allocation for all the newly introduced variables.
\end{itemize}

The key insight into why this leads to significant memory savings, is the fact
that in all polyhedral programs that we encountered, truly affine dependences
are almost always \emph{rank deficient}, i.e., are many-to-one mappings from
the consumer index points to the producers.  The only exceptions are either
pathological programs, or programs that do multi-dimensional data
reorganizations via bijections (e.g., matrix transpose, tensor permutations,
etc.) where here is no scope nor need to reduce the total memory footprint.
As a result, $f(D_e)$ is almost always a lower dimensional polyhedron, and
requires significantly less memory, even when stored supposedly inefficiently.

\subsection{Related Work}

Memory allocation for polyhedral programs is a well studied problem for almost
two decades.  DeGreef and Cathoor~\cite{degreef-memory97} tackled the problem
of sharing the memory across multiple arrays in the program.the so called
inet-array memory reuse problem, and proposed an ILP based solution.  Wilde
and Rajopadhye, in dealing with an intrinsically memory-inefficient functional
language Alpha~\cite{mauras1989thesis} (one can think of this as a program after
full expansion) first addressed the memory reuse for points of an iteration
space~\cite{sanjay-europar96}.  They gave necessary and sufficient conditions
for the legality of a memory allocation fucntion, which they allowed to be
``in any direction.''  but they did not provide any insight into how to choose
the mapping.  Lefebvre and Feautrier~\cite{lefebvre-feautrier-pc98} on the
other hand, considered only canonic projections, combined with a modulo
factor, but showed how to choose the mapping optimally.  Later, Quiller\'e and
Rajopadhye~\cite{sanjay-toplas00} revisited multiprojections, extended them to
quasi-affine functions, and proved a tight bound on the number of dimensions
of reuse.  They also showed that cananic projections with modulo factors was
sometimes a constant factor better, and sometimes a constant factor worse.
Darte at al.~\cite{darte-lattice05} took a fresh and elegant approach to the
problem, and formulated the conditions for legal memory allocations by
defining the \emph{conflict set}.  This led to techniques for choosing
provably optimal memory allocations, initially for non-parameterized iteration
spaces, and recently in the context of FPGA acelerators, for parametrically
tiled spaces~\cite{darte2014parametric, darte2016extended}.  Vasilache et
al.~\cite{vasilache-impact12} developed a tool to combine the scheduling and
limiting memory expansion using an ILP formulation, implemented in the
R-Stream compiler.  Recently, Bhaskaracharya et
al.~\cite{bhaskaracharya-toplas16} developed methods to optimally choose
quasi-affine memory allocations, and showed how they are beneficial for tiled
codes, especialy with live-out data.  Furthermore, they also
showed~\cite{bhaskaracharya-popl16} how to combine iner-and intra array reuse
in a unifying framework.

The other \emph{schedule independent} memory allocation was pioneered by
Strout et al.~\cite{strout-etal-asplos98}.  Here, the memory allocation is
chosen based only on the dependences, and is guaranteed to be legal,
regardless of the schedule.  This problem is solved when the program has
\emph{uniform dependences}, i.e., when each dependence can be descibed by a
\emph{constant vector}, and for some simple extensions of
this~\cite{strout-etal-asplos98, sanjay-memory-2011}.

Thies at al.~\cite{thies-pldi02} have also formulated the problem of
simultaneously choosing the schedule and memory allocation as a combined
optimization problem.
  
  
    	



\end{document}